\documentstyle[aps,prl,preprint,epsfig]{revtex}
\def\gappeq{\mathrel{\rlap{\raise.5ex\hbox{$>$}}
{\lower.5ex\hbox{$\sim$}}}}
\def\lappeq{\mathrel{\rlap{\raise.5ex\hbox{$<$}}
{\lower.5ex\hbox{$\sim$}}}}
\begin{document}
\topmargin -1.0cm
\oddsidemargin -0.8cm
\evensidemargin -0.8cm
\pagestyle{empty}
\begin{flushright}
CERN-TH/99-341
\end{flushright}
\vspace*{5mm}
\begin{center}
{\Large\bf Gauge bosons in a five-dimensional theory
}\\
{\Large\bf with localized gravity}\\
\vspace{2cm}
{\large\bf Alex Pomarol \footnote{On leave of absence from
IFAE, Universitat Aut{\`o}noma de Barcelona, 
E-08193 Bellaterra, Barcelona.}}\\
\vspace{.6cm}
{\it {Theory Division, CERN}\\
{CH-1211 Geneva 23, Switzerland}\\}
\end{center}
\begin{abstract}
We consider the possibility of 
gauge bosons living in the recently proposed 
five-dimensional theory with localized gravity.
We study the mass spectrum of the
Kaluza-Klein (KK) excitations of the gauge
fields  and calculate their couplings to the boundaries
of the fifth dimension.
We find a different  behaviour from the case of the graviton.
In particular, we find that the massless mode is not localized
in the extra dimension and that the KK excitations have 
sizeable couplings to the two boundaries.
We also discuss  possible phenomenological implications
 for the case of the standard model  gauge bosons.
\end{abstract}
\vfill
\begin{flushleft}
CERN-TH/99-341\\
November 1999
\end{flushleft}
\eject
\pagestyle{empty}
\setcounter{page}{1}
\setcounter{footnote}{0}
\pagestyle{plain}


\section{Introduction}

It has been recently realized that extra dimensions can play a crucial
role in the hierarchy problem. 
A first example was given in Ref.~\cite{add} where
the weakness of the gravitational
interaction versus the gauge interactions was obtained by
allowing the graviton to propagate in some extra (and very large) 
dimensions.

An alternative  scenario has been considered in Refs.~\cite{rs,gob}.
They assumed an extra dimension
 compactified on a  segment $S^1/Z_2$
with two 4D boundaries.
By adjusting the cosmological constant of the 
boundaries with  the one in the bulk, they found a solution
to the 5D metric of the form
\begin{equation}
  \label{metric}
ds^2=e^{-2kR\phi}
\eta_{\mu\nu}dx^\mu dx^\nu-R^2d\phi^2\, ,
\end{equation}
where $\mu=1,..,4$, and $R$ is the radius of the compact fifth-dimension
parametrized by $\phi$ with $0\leq \phi\leq\pi$. 
$k$ is a parameter related to the cosmological constant \cite{rs}.
The metric (\ref{metric}) corresponds to a slice of $AdS_5$
between the
two boundaries  located at $\phi=0$ and $\phi=\pi$.
After reducing to 4D, one finds that the 
effective 4D Planck scale  is given by $M^2_P=M^3_5(1-e^{-2kR\pi})/k$
where $M_5$ is the 5D Planck mass, the fundamental scale
in the theory. 
Masses on the  4D boundary at $\phi=\pi$, however, 
appear to be  reduced 
by the  exponential factor $e^{-kR\pi}$ 
of the induced metric on the boundary.
Therefore, for $kR\simeq 12$ and $k\sim M_5$, 
one can generate two scales 
for the fields living 
 on the boundary at $\phi=\pi$: the Planck scale
$M_P\simeq M_5$ and  the electroweak scale $M_Pe^{-kR\pi}\sim$ TeV.
Fields living in the 5D metric 
(\ref{metric}) have a very peculiar 
Kaluza-Klein (KK) decomposition. 
The  masses of the KK are exponentially suppressed
 $M_Pe^{-kR\pi}\sim$ TeV and therefore are very light
even though the radius of compactification is (approximately)
of Planckian size.
These KK modes have extensively been studied for the case of the graviton
\cite{rs,others} and a scalar field \cite{wise}.

In this letter we want to consider the case of a  gauge boson
propagating in the $AdS_5$ slice described above.
We will study the KK decomposition of the gauge boson field
and analyze its phenomenological implications.
We will show that, unlike the graviton or the scalar,  the zero mode 
of the gauge boson is not localized in the 5D theory.
Similar to the case of a flat extra dimension, this massless gauge boson
spreads over the extra dimension.
Also their corresponding  KK excitations behave differently from those
of  the graviton or scalar. 
Although
the KK masses are also exponentially suppressed, their couplings to
the fields on the boundary  at $\phi=0$ are not suppressed.
They can therefore lead to sizeable effects.  
On the boundary at  $\phi=\pi$, we will show that the 
coupling of the KK excitations
 depends linearly on $\sqrt{kR}$ and therefore the  KK become
strongly coupled for $kR\sim 12$.

\section{KK decomposition of a 5D gauge boson}

The equation of motion of a U(1) gauge boson $A_N$ of mass $M$
in a curved space is given by
\begin{equation}
  \label{motion}
\frac{1}{\sqrt{g}}\partial_M\left(
\sqrt{g}\, g^{MN} g^{RS} 
F_{NS}\right)-M^2g^{RS}A_S=0\, ,
\end{equation}
where $g=Det(g_{MN})$, $F_{MN}$ is the gauge-field strength
and we denote with
 capital Latin letters the 5D coordinate 
$M=(\mu,\phi)$.
For the metric of eq.~(\ref{metric}),   eq.~(\ref{motion})
leads to \footnote{For the massless case $M=0$, we  consider the 
gauge $\partial_\mu A^\mu=0$ and $A_\phi=0$.}
\begin{equation}
  \label{motion2}
 \left[\eta^{\rho\nu}\partial_\rho\partial_\nu+\frac{1}{R^2}\partial_\phi
e^{-2kR\phi}\partial_\phi
-e^{-2kR\phi} M^2\right]A_\mu=0\, .
\end{equation}
We can decompose the 5D field as
\begin{equation}
  \label{kk}
A_\mu(x^\mu,\phi)=\sum_n A^{(n)}_\mu(x^\mu)\frac{f_n(\phi)}{\sqrt{R}}\, ,  
\end{equation}
where $f_n$ satisfies
\begin{equation}
  \label{kkequation}
-\left[\frac{1}{R^2}\partial_\phi
e^{-2kR\phi}\partial_\phi+
e^{-2kR\phi} 
M^2\right]f_n=m^2_n f_n
\, .
\end{equation}
The  $m_n$ correspond to the  masses of the KK excitations $A^{(n)}_\mu$.
These masses and the corresponding eigenfunctions
 $f_n$ can be found solving
eq~(\ref{kkequation}) with the boundary condition imposed by the orbifold
$S^1/Z_2$
(this requires $f_n(\phi)=f_n(-\phi)$, and
that $f_n$ and its derivative are continuous at $\phi=0$ and $\phi=\pi$.).
We find
\begin{equation}
  \label{solution}
f_n=\frac{e^{kR\phi}}{N_n}\left[J_\alpha(\frac{m_n}{k}e^{kR\phi})
+b_{\alpha}(m_n)\, Y_\alpha(\frac{m_n}{k}e^{kR\phi})\right]\, ,
\end{equation}
where $J_\alpha$ and $Y_\alpha$ are the Bessel function of order $\alpha$
with
$\alpha=\sqrt{1+M^2/k^2}$ and $b_{\alpha}(m_n)$ are coefficients
that fulfil
\begin{equation}
  \label{b1}
  b_{\alpha}(m_n)
=-\frac{J_\alpha(\frac{m_n}{k})+\frac{m_n}{k}
J^\prime_\alpha(\frac{m_n}{k})}
{Y_\alpha(\frac{m_n}{k})+\frac{m_n}{k}
Y^\prime_\alpha(\frac{m_n}{k})}\, ,
\end{equation}
and
\begin{equation}
  \label{b2}
  b_{\alpha}(m_n)=  b_{\alpha}(m_n e^{kR\pi})\, .
\end{equation}
$N_n$ is a normalization factor defined such that
\begin{equation}
  \label{norm}
\int^\pi_0f^2_nd\phi=1\, .  
\end{equation}
For $kR\gg 1$, we have
\begin{equation}
  \label{N_n}
 N^2_n\simeq \frac{e^{2kR\pi}}{2kR}J^2_\alpha(\frac{m_n}{k}e^{kR\pi})\simeq
\frac{e^{kR\pi}}{\pi R m_n}\, .
\end{equation}
From the condition (\ref{b2}), one can obtain 
the eigenmasses $m_n$.
Let us consider separately the case of a 5D massless vector
boson $(M=0)$ and a massive one
($M\not=0$).

\subsection{Massless 5D gauge boson ($M=0$ case)}

In this case 
the lowest mode is a massless state $A^{(0)}_\mu$ with 
\begin{equation}
  \label{zeromode}
f_0=\frac{1}{\sqrt{\pi}}\, .  
\end{equation}
This is a constant mode. Therefore, unlike the graviton case,
it is not localized in the fifth dimension. 
It
couples with equal strength to the two boundaries
$g=g_5/\sqrt{\pi R}$ where $g_5$ is the 5D gauge coupling.
Since we are assuming that $R$ is close to $k\sim M_P$, 
this massless mode has a coupling 
constant of order unity. 

To obtain the massive modes, we first notice that 
$\alpha=1$ and   eq.~(\ref{b1}) reduces to
\begin{equation}
  \label{b3}
  b_{1}(m_n)
=-\frac{J_0(\frac{m_n}{k})}{Y_0(\frac{m_n}{k})}\, .
\end{equation}
For $m_n\ll k$, we have
\begin{equation}
  \label{b3app}
  b_{1}(m_n)
\simeq-\frac{\pi/2}{\ln(m_n/2k)+\gamma}\, ,
\end{equation}
where $\gamma$ is the Euler constant.
The KK masses can be derived from
eq.~(\ref{b2}) that now takes an approximate form:
\begin{equation}
  \label{b4}
J_0(\frac{m_n}{k}e^{kR\pi})\simeq 0\, .
\end{equation}
The solution are for $n=1,2,...$
\begin{equation}
  \label{b4sol}
m_n\simeq (n\pi-0.7) k\, e^{-kR\pi}\, .
\end{equation}
We see then that the masses of the KK modes are exponentially
suppressed and thus $m_n\sim$ TeV (for the case  $kR\sim 12$
that we are considering).
To know how they couple to the boundaries, we must 
look at their eigenfunctions $f_n$ near the boundaries.
For the boundary at $\phi=0$,
the $f _n$ are well approximated 
by the second term in eq.~(\ref{solution}):
\begin{equation}
  \label{branecoupling}
f_n(\phi=0)\simeq \frac{1}{N_n}
b_1(m_n) 
Y_1(\frac{m_n}{k})
\simeq -\frac{1}{N_n}b_1(m_n) \frac{2k}{\pi m_n}\, .
\end{equation}
From 
eqs.~(\ref{N_n}) and (\ref{b3app})-(\ref{branecoupling}),
we obtain that the coupling of the KK modes to the fields
on the boundary at $\phi=0$ are
\begin{equation}
  \label{branecouplingg}
g^{(n)}=g\sqrt{\pi}f_n(\phi=0)\simeq g
\sqrt{\frac{\pi kR}{n-0.2}}\left(\ln\frac{n\pi-0.7}{2}-kR\pi
+\gamma\right)^{-1}\, .  
\end{equation}
This coupling is of order $g$.
For the $n=1$ mode we get $g^{(1)}\simeq 0.2\, g$ (for $kR\sim 12$).
Note that this is  different from the graviton case, 
where the coupling
of the light KK excitations to the boundary at $\phi=0$
is exponentially suppressed \cite{rs,others}.
The coupling of the KK gauge bosons to the fields on the 
boundary at $\phi=\pi$  is  dominated by the 
first term of eq.~(\ref{solution}) that leads
to  $g^{(n)}\simeq g\sqrt{2\pi kR}$. 
Since it grows with the radius $R$,
we  can obtain an  upper bound on $R$ by imposing
that the theory is perturbative, i.e. $g^{(n)\, 2}/4\pi<1$.
We get
\begin{equation}
\label{bound}
R\lappeq \frac{2}{kg^2}\, .
\end{equation}
This bound is quite strong 
and seems to disfavor this type of scenarios with $kR\sim 12$.

\subsection{Massive 5D gauge boson ($M\not=0$ case)}

For a massive 5D vector boson, the results are slightly
 different.
The eigenmasses are, approximately, given by
\begin{equation}
  \label{b4app2}
J_\alpha(\frac{m_n}{k}e^{kR\pi})+
\frac{m_n}{k}e^{kR\pi}
J^\prime_\alpha(\frac{m_n}{k}e^{kR\pi})
\simeq 0\, ,
\end{equation}
that leads again  to exponentially suppressed masses, 
$m_n\sim k e^{-kR\pi}\sim $ TeV.
An important difference from the $M=0$ case is the coupling of
these modes to the $\phi=0$ boundary.
We have now
\begin{equation}
  \label{branecouplingg2}
g^{(n)}\simeq -g\sqrt{\pi}
\frac{2\alpha}
{N_n(1-\alpha)\Gamma(1+\alpha)}
\left(\frac{m_n}{2k}\right)^{\alpha}\, .
\end{equation}
We see that now the coupling to the boundary at $\phi=0$ is very small,
due not only to the exponential suppression (coming from $N_n$) but also
to  the  $m_n/k$ suppression.
The coupling to the boundary fields at $\phi=\pi$ is found to be
the same as for the massless case,
$g^{(n)}\simeq g\sqrt{2\pi kR}$.  

Finally it is interesting to point out what has  
happened to the massless mode that appeared in the $M=0$ case.
This mode is still present in the theory but its mass has become
of order $M$.  

\section{Phenomenological speculations}

The above analysis 
indicates that massless gauge bosons living in a  5D bulk 
with  the metric (\ref{metric})
and  $kR\simeq 12$ 
have very light KK excitations (of order TeV).
They   couple 
with a strength  $\sim 0.2\,  g/\sqrt{n}$ 
to the fields on the boundary at $\phi=0$,
but are strongly coupled ($g^{(n)}\simeq 8\, g$)
to the fields on the boundary at $\phi=\pi$.
Massive 5D gauge bosons also have TeV KK excitations, but they 
couple very weakly to the boundary fields at $\phi=0$.

In this section we want to consider the possibility
of having 
the standard model (SM) fermions localized on the 4D boundary at $\phi=0$
with  the SM gauge bosons living in the 5D bulk \footnote{
In this scenario supersymmetry  could be needed in order to preserve
the Higgs mass smaller than $M_P$. 
Our motivation here will not be the hierarchy problem, but
to analyze the distinctive
phenomenological implications of this scenario.}.
In this case
the KK gauge bosons will have non-negligible  couplings to the SM fermions.
This looks quite surprising,  since even though 
the cutoff of the theory for the fields on the boundary is
$M_P$, the KK excitations can affect low-energy processes.
Bounds on $R$ can  be obtained, for example, 
from electroweak high-precision measurements as in Ref.~\cite{em}.
We get in this case
a bound on the first KK mass $m_1>$ 300--500 GeV.
Deviations in four-fermion interactions can also be  used in future
colliders to test  these scenarios
similarly to the case
of a large (TeV) extra dimension.
We must stress, however,  that in the scenarios considered here
there is an important difference 
from the case of a
large TeV-dimension.
In this latter case, the strength of the interaction
of two fermions on the boundary with  a   gauge boson in the bulk 
grows, at the classical level,  with the square-root of the energy 
(since for energies above the first KK mass, the  number of KK
that mediates the interaction  increases with the energy).
This means that the theory becomes  strongly coupled   above the first KK
mass,
and we loose perturbativity (new physics must appear).
Here, however, the coupling constant of the $n$th KK excitation
goes as $g/\sqrt{n}$ (see eq.~(\ref{branecouplingg})),
and therefore the strength of the interaction
remains constant. At the quantum level, however,
the gauge boson propagator  is sensitive to the KK excitations
and 
it is not clear if the theory will remain perturbative
or it will become strongly coupled, and therefore inconsistent.
Further studies along these lines are clearly needed.

If this theory is embedded 
at high energies $\sim M_P$
into  a GUT group (such as SU(5)),
one could expect  that large  proton-decay operators will be
 induced, since
the KK excitations of the 
GUT gauge bosons will be  very light (they will have TeV masses). 
Nevertheless, we have  seen that  the coupling 
of the KK excitations of a 
massive 5D gauge boson  
to the fermions on  the $\phi=0$ 
boundary 
is small ---
see eq.~(\ref{branecouplingg2}). 
Even if we sum over the full KK tower, we get 
that any
four-fermion operator induced by a massive 5D gauge boson field
is strongly suppressed $\sim 1/k^2$  (for a 5D gauge-boson mass
 $M$ of order $k$).
The coupling of the SM gauge bosons to the 
KK excitations of the 
GUT fields is, however,  
not suppressed, since massless gauge-modes can propagate in the 
extra dimension. Consequently, 
KK excitations of the GUT fields could be produced in pairs by 
a Drell-Yan process.
It is interesting  to recall that, as we said at the end of section~2,
the  GUT-partners of the  massless SM gauge bosons
are  modes with masses $M\sim k$
(these are modes that would tend to a  
massless and $\phi$-independent 
mode if the VEV that breaks the GUT tends to zero).
These modes   seems therefore that will induce, 
at the one-loop level,
a logarithmic
dependence of the gauge coupling with the high-energy scale ($\sim M_P$)
as in the case of a 4D theory.

Summarizing, we have studied the behaviour of a 
U(1) gauge boson field propagating in a slice of $AdS_5$.
We have analyzed its KK decomposition 
and find that there are very light KK excitations
that couple with strength $g$ to the fermions on the boundaries.
The  theory, on the boundary at $\phi=0$,
seems to have an interesting behaviour. 
A priori the 
cutoff
 scale of the theory for  fermions at $\phi=0$ is  $M_P$.
Nevertheless, it is possible from that boundary
 to ``see''  physics at $M_P$ since
the KK modes are light and couple to the fermions with non-negligible
couplings.
Further studies, however, are needed in order to assure
 the consistency of the theory at the quantum level.

We thank Jaume Garriga, 
Tony Gherghetta and Daniel Waldram
for  discussions.

 \vspace{1cm}
{\bf Note added:} While writing this paper,  it appeared Ref.~\cite{new}
considering also gauge bosons propagating in the metric (\ref{metric}).

\end{document}